\documentclass{article}
\usepackage[utf8]{inputenc}
\usepackage{graphicx}
\usepackage{xcolor}
\usepackage{hyperref}
\usepackage{url}

\title{The Virtual Emotion Loop: Towards Emotion-Driven Product Design via Virtual Reality}
\author{\small Davide Andreoletti, Luca Luceri, Achille Peternier, Tiziano Leidi, Silvia Giordano 
	\\ \footnotesize University of Applied Sciences and Arts of Southern Switzerland (SUPSI)}

\begin{document}

\maketitle

\begin{abstract}

Emotions play a significant role in the interaction between products and users. However, it is still not very well understood how users' emotions can be incorporated in product design. We argue that this gap is due to a lack of a methodological and technological framework for an effective investigation of the elicitation conditions of emotions and of the corresponding emotional response of the users. Indeed, emotions are either disregarded or not systematically considered in current design approaches. For example, the effectiveness of emotion elicitation conditions embodied into the product is generally validated by assessing users' emotional response through ineffective means (e.g., surveys and interviews). In our view, emotion-driven design should encompass a thorough assessment of users' emotional reactions in relation to certain elicitation conditions. In this paper, we argue that Virtual Reality (VR) is the most suitable mean to perform this investigation, and we propose a novel methodological framework, referred to as the Virtual-Reality-Based Emotion-Elicitation-and-Recognition loop (VEE-loop), that can be exploited to realize it. Specifically, the VEE-loop consists in a continuous monitoring of users' emotions, which are then provided to product designers as an implicit users' feedback. This information is used to dynamically change the content of VR environment, and the process is iterated until the desired affective state is solicited. The main applications that we envision are in product design (i.e., to early validate the effectiveness of some emotion elicitation conditions) and digital product delivery, e.g., distance learning can be adapted in real-time according to the user's emotional status.

%e discuss issues and opportunities of this VEE-loop, and we also present potential applications of the VEE-loop in research and in various application areas.  

\end{abstract}

\section{Introduction}

The capability of a product (being it tangible or not, e.g., a service) to engage its users at the emotional level is considered by many the main factor behind its success \cite{alaniz2019emotional}. In this respect, the authors of Ref. \cite{zaltman2003subconscious} claim that up to $95\%$ of our buying decisions are not driven by rational arguments. %Moreover, it has been shown that consistency of users' emotional responses across various phases of the user-product relation (e.g., purchase and use) ensures long-term brand loyalty \cite{wrigley2019affected}. 
Despite this, the fulfillment of functional requirements is generally the only objective in product %s and in services 
design, with emotional aspects being often underestimated, if not even totally disregarded \cite{wrigley2019affected}. While this contradiction is commonly interpreted as a phenomenon of cultural inertia \cite{wrigley2019affected} (argument that we also support), in this paper we give the following additional interpretation: there is no methodological and technological framework that helps designers to systematically experiment emotion elicitation conditions, as well as to assess the consequent users' emotions. %Such framework could be used, for example, to perform the validation of emotional requirements, i.e., to check the consistency between intended and perceived emotions. 

\textit{Emotion-driven design} is the set of processes and methods used for developing products with the specific aim of evoking certain emotional responses. Following this paradigm, designers consider the final users' emotions since the very early stages of development, and not, as often done, by leaving them as an afterthought. Existing approaches aimed to foster emotion-driven design (e.g., \cite{alaniz2019emotional}) are based on the idea that designers should familiarize with all the aspects of emotions, ranging from their definition and elicitation conditions (e.g., in terms of products' sensory characteristics) and manifestations on people. Following a similar line of reasoning, in this paper we argue that emotion-driven design should be characterized by i) a systematic experimentation of various emotion elicitation conditions (e.g., the sensory qualities of the product and context of its usage) and by ii) the reliable measurement of users' emotional response, i.e., emotion recognition. Then, we also argue that emotional elicitation and emotion recognition should be iterated in a continuous loop until the emotion elicitation conditions can evoke the emotion intended by the designer.  

Following the scheme envisioned in this paper, a designer makes her stylistic choice with the intention to evoke some emotional response (i.e., emotion elicitation phase); users' emotions are qualitatively measured (i.e., emotion recognition phase); the designer observes the emotions actually perceived by users and gains relevant insights to adjust her choices accordingly (i.e., loop phase). We are aware that the emotional reaction might be significantly affected by the number of iterations. Just as an example, boredom might be evoked more frequently after having done several experiments, regardless of the current emotion elicitation conditions. We discuss this and similar issues in Section \ref{sec:challenges}. Please note that the proposed scheme can be implemented only in the design phase (e.g., to validate the hypothesis that some feature triggers a specific emotional reaction) or, whenever possible, in the delivery phase as well (e.g., by dynamically adapting the characteristic of a service in real-time). While the proposed scheme can, in principle, be implemented using the most varied approaches, we believe that Virtual Reality (VR) provides the perfect controlled environment to turn our vision into a practical design instrument. Therefore, we refer to the proposed framework as the VEE-loop, i.e., the Virtual-Reality-Based Emotion-Elicitation-and-Recognition loop.

We identify several factors that, in our view, make VR the most suitable technology to implement the proposed scheme. First, among all the existing digital technologies, VR is the one that guarantees the most tangible experience across different domains. In fact, VR allows users to feel a sense of presence that makes emotion elicitation conditions quite similar to a real scenario \cite{riva2007affective}, with the remarkable advantage of also enabling a flexible modification of the virtual scene experienced by the user, i.e., the Virtual Environment (VE). In addition, VR allows gathering a set of users-related data from which their emotions can be inferred (e.g., bio-feedback and behaviors). Note that a number of bio-feedback signals can be gathered using the Head Mounted Displays (HMDs) used by VR system, as well as with other external devices (e.g., wearables). 

The VEE-loop has the potential to benefit a high number of application areas. For example, it can be used as a tool to perform a validation of the capability of a product to trigger specific emotions, before its actual production. Indeed, designers could obtain an emotional feedback from potential customers and tune the design accordingly. Given that this feedback is obtained before the actual product development, the risk of designing unsuccessful products is significantly reduced. In this respect, the immersion level provided by VR guarantees a higher fidelity of this emotional reaction with respect to other methods, while its flexibility allows testing a high number of products' characteristics. The VEE-loop can also improve services in which the knowledge of users' affective states is highly beneficial, but unavailable for some reason (e.g., due to physical distancing measures imposed to handle the Covid-19 pandemic). In remote learning, for example, the emotional states of students can be monitored, and the virtual lecture dynamically changed (e.g., to induce calm in students or to draw their attention) \cite{ortigosa2014sentiment}. 

This paper is structured as follows. In Section \ref{sec:towards_eds} we elaborate on the importance of emotions in user-product interaction, we present the characteristics of emotion-driven product design and we motivate the use of VR as enabling technology to implement it. Then, in Section \ref{sec:vee-loop} we describe the VEE-loop in detail, while in Section \ref{sec:soa} we show how the VEE-loop advances existing approaches. Section \ref{sec:impact_application} is devoted to the presentation of the impact and of the application areas of the VEE-loop. Finally, in Section \ref{sec:challenges} we discuss opportunities and challenges derived from the use of the VEE-loop, also adding some concluding remarks. 

\section{Towards VR-based Emotion-Driven Product Design}\label{sec:towards_eds}

This Section starts by elaborating on the importance of emotions in users-product interaction. Then, it introduces the paradigm of emotion-driven product design, highlighting its main characteristics. Finally, we motivate the use of VR as an enabling technology towards the realization of this paradigm.

\subsection{Emotions in users-product interaction}

Emotions are present in almost all human experiences, including the interaction between users and products. Based on the findings of previous works, the authors of Ref. \cite{alaniz2019emotional} identify three main products' characteristics that evoke emotions on their users, namely \textit{appearance}, \textit{functionality} and \textit{symbolic meaning}. As for the appearance, it is acknowledged that sensory qualities (e.g., shape and color) are associated with different emotional experiences. For instance, warm colors are generally chosen to increase the arousal levels of evoked emotions \cite{liu2013eeg}. In general, the functionalities of a product elicit positive emotions if they fulfill the needs of the users, and negative otherwise. For instance, a product that improves a situation that is perceived as frustrating and limiting (e.g., by enabling to gain space in small environments) is likely to evoke positive emotions. On the contrary, a product that is cumbersome and reduces the available space likely leads to frustration and annoyance. Then, the symbolic meaning of a product refers to its connection with a broader scheme of beliefs and values. In relation to the symbolic meaning of a product, the appraisal theory \cite{desmet2002designing} states that emotions are triggered by the foretaste that users have when they evaluate the role of the product in their lives. For instance, a treadmill can evoke positive emotions in those who see it as a mean to get fitter, but negative ones in those worried by strain. Another example is the symbolic value given by the affinity of a product with a certain idea (e.g., a flag that represents a certain political view). 

The importance of the symbolic meaning of products in evoking emotions is well expressed in the famous quote stated by Simon Sinek in one of the most viewed TED talk ever\footnote{\url{https://www.ted.com/talks/simon_sinek_how_great_leaders_inspire_action?language=en}}: \textit{people don't buy what you do, they buy why you do it} \cite{sinek2009start}. Indeed, the meaning that a person ascribes to a product is strongly correlated with her inner values, and their affinity with a company's mission and concerns. In relation to this, the \textit{law of concern} formulated in \cite{frijda1986emotions} affirms that every emotion hides a personal concern and a disposition to prefer particular states of the world. This fact has led the authors of Ref. \cite{desmet2016emotion} to describe emotions as gateways to what people really care for, and entry points to uncover their underlying concerns. Along similar lines, the authors of Ref. \cite{wrigley2019affected} argue that understanding users at the emotional level allows having a deeper comprehension of their values, which is crucial to produce radical product innovations, while the sole understanding of users' functional needs yields only superficial and slight product modifications. Moreover, products capable to emotionally engage their users foster creative and innovative thinking \cite{desmet2016emotion}, and benefit well-being \cite{kim2021designing}. Therefore, the capability of understanding and engaging users at the emotional level is crucial to design products that are appreciated and guarantee loyalty of customers in the long term (in this respect, note also that a clear and tight connection between a product and a specific emotion reinforces brand identification \cite{wrigley2019affected}).

\subsection{Characteristics of emotion-driven product design}

Given the major role of emotions in the relation between users and products, \textbf{emotion-driven design}, i.e, the realization of products with the deliberate intention to evoke specific emotions \cite{desmet2016emotion}, is rapidly becoming an important research area. In particular, several frameworks have been recently proposed to help designers in the creation of products with emotional intentions. These frameworks (e.g., the Emotion-Driven Innovation paradigm \cite{alaniz2019emotional}) share the idea that designers should be supported in the acquisition and practical exploitation of a solid \textit{emotion knowledge}, which is defined in Ref. \cite{desmet2016emotion} as the explicit understanding of the physical manifestations of emotions and of their eliciting conditions. At the moment, these frameworks are still not very diffused \cite{kim2021designing}; arguably, this limited diffusion is mainly due to the lack of a technological layer to facilitate the study of emotion elicitation and recognition. In this paper, we claim that VR represents the most powerful medium to invert this tendency and consolidate the practice of emotion-driven design. In the following, we express our view on the characteristics that a framework for emotion-driven design should have, both concerning the study of the conditions of emotion elicitation and the recognition of emotions from their manifestation. Then, in subsection \ref{sec:VR}, we provide arguments that support the idea of using VR as the basis to implement such a framework. 

\subsubsection{Emotion Recognition}

The capability to disambiguate between different emotions (e.g., to understand the difference between frustration and annoyance) has been defined in Ref. \cite{desmet2016emotion} as emotional granularity and is regarded in Ref. \cite{yoon2016feeling} as a core advantage for the realization of emotion-driven products. Indeed, it is essential that designers understand the nuances of emotions beyond the simple positive versus negative distinction \cite{kim2021designing} (in this respect, just think that consumers may experience $25$ different positive emotions when interacting with a product \cite{desmet2016emotion}). 

The most straightforward approach to understand users' emotions is by direct communication. However, people are generally not aware of their emotions, nor they can properly verbalize and communicate them. Hence, traditional investigation tools (e.g., surveys and interviews) cannot effectively capture users' emotions. Moreover, tools based on self-reports require users to interrupt their activity, which in turn may hinder the validity of their records \cite{desmet2016emotion}. Therefore, emotion recognition techniques, i.e, the qualitative measurement of emotions from their manifestations, seems to be the most viable alternative. 

Affective computing refers to a set of technologies and strategies developed to automatize emotion recognition, generally exploiting machine learning algorithms that infer the emotions a person most likely perceives from her bio-feedback (e.g., facial expression, blood pressure, movements, etc.). The fact that only bio-feedback are considered is, in our view, a severe limitation of the traditional approach. For instance, behaviors of users, which are actually part of the manifestations of emotions \cite{kim2021designing}, are currently disregarded. Indeed, emotions are complex phenomena that are better understood if studied holistically \cite{kim2021designing}. In user-product interaction, this holistic investigation would require, for example, the correlation of the sensory and symbolic characteristics of the product with the bio-feedback of the users, as well as with her behaviors (e.g., which action users take after using a product) and with the context in which the product is used \cite{desmet2016emotion}. In particular, the possibility to correlate users' emotions with contextual factors has been considered in Ref. \cite{kim2021designing} as a way to better understand their concerns and inner values. Therefore, we believe that a framework for emotion-driven design should allow probing users' emotions considering as many aspects of their manifestations as possible (e.g., bio-feedback, context, behaviour, etc.). 

\subsubsection{Emotion Elicitation}

The task of identifying the conditions that elicit certain emotions is inherently challenging. Indeed, while it is acknowledged that certain products' characteristics induce similar emotional reactions on the majority of their users \cite{liu2013eeg}, emotions are generally subjective experiences. In other words, the relation between certain types of stimuli and emotions is neither deterministic nor constant, as it can change over time even for the same person \cite{desmet2016emotion}. In addition, elicitation conditions are extremely complex, as they depend on the interaction of various factors, such as product's characteristics and context of usage.

In light of this, we believe that a framework for emotion-driven design should favour the validation of emotion elicitation conditions both on a single-user and on a large-scale basis. As for the former, this framework would help designers to understand a customer at the emotional level, and in turn uncover her latent concerns and desires \cite{wrigley2019affected}. In addition, this framework would enable the design of products performed cooperatively by designers and customers (note that product co-design is considered more attractive than designer-only and customer-only design \cite{wrigley2019affected}). As for the latter, the framework should allow to easily validate the effectiveness of different combinations of elicitation conditions (e.g., sensory qualities of the product and environment in which the product is used) in evoking certain emotions.

\subsubsection{Loop of Emotion Recognition and Elicitation}\label{sec:loop}

\begin{figure}[t!]
    \centering
    \includegraphics[width=\linewidth]{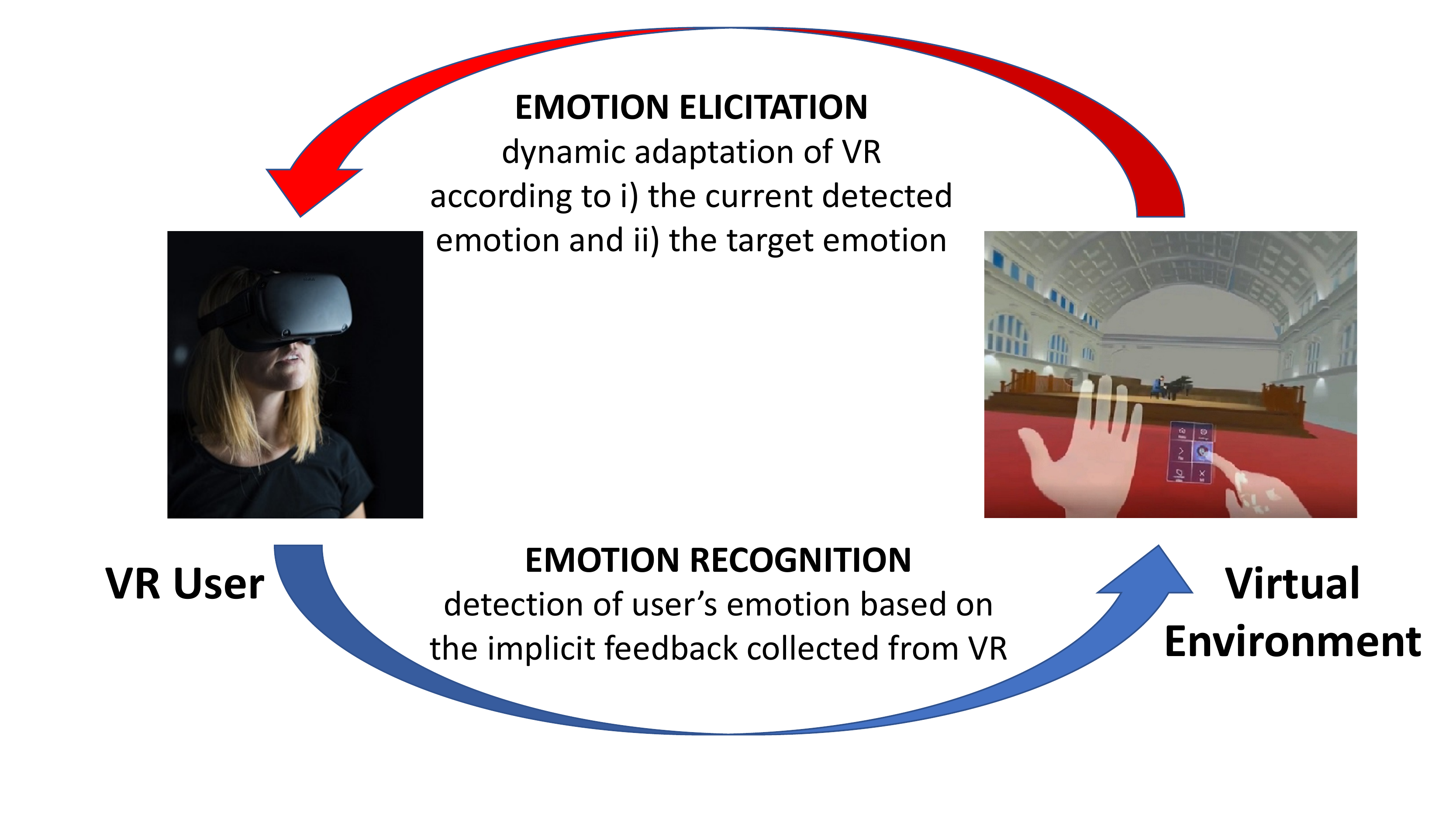}
    \caption{High-Level Representation of the VEE loop}\label{fig:veexxloop}
\end{figure}

Finally, our envisioned framework for emotion-driven product design is based on a tight connection between emotion recognition and emotion elicitation. In fact, users' emotions should be continuously tracked and provided to designers as a feedback of the suitability of the chosen elicitation conditions. Elicitation conditions can then be modified accordingly, in a more informed manner. This operation can then be repeated until the expected emotional reaction is achieved, which gives rise to the \textbf{Emotion-Recognition-Emotion-Elicitation Loop}. This scheme can be applied, for instance, to validate the effectiveness of experimental emotion elicitation conditions, as well as to better understand the factors that triggered some particular emotions. In the following subsection, we provide arguments that support our choice of VR as the most suitable candidate technology to implement this scheme, which we then refer to as the \textbf{VR-based Emotion-Recognition-Emotion-Elicitation Loop}, or simply the VEE-loop. A representation of the VEE-loop is depitected in Fig. \ref{fig:veexxloop}.

\subsection{VR as enabling technology}\label{sec:VR}

%\textcolor{blue}{here we say that VR gives the possibility to implement such a service/product because of ... Elaborate on the flexibility, non-invasive nature, tangibility, etc etc }

For various reasons, VR is the most natural, direct, and suitable technology to implement the framework described so far. First, pure VR (i.e., a user interacting with entirely synthetic, computer-generated VEs) allows creating completely modifiable, dynamic experiences. Unlike augmented and mixed reality, which is limited and linked to the surrounding physical elements, pure VR can be easily distributed online, experienced everywhere, replayed at will, and its content regularly updated. The immersion provided by VR also amplifies emotional reactions \cite{diemer2015impactPerception, morales2019realVsVRemotion,riva2007affective}, which helps both in the emotion recognition and elicitation phases compared to other less effective means to put a user in a given simulated situation. In addition, the retention rate of learning and training provided via VR is increased when compared to more conventional media \cite{babu2018vrLearningRate}. The recent revival of VR also provides a much lower entry-point to the technology, which is now considered a commodity, off-the-shelf option no longer limited to research laboratories or professional contexts. Thanks to this evolution, which also significantly increased the quality of modern VR compared to the state of the art of just few years ago, a significantly larger user-base can now be targeted by VR-enabled solutions. 

Then, modern VR equipment already embeds sensors that are critical for inferring the user's emotional state (and its evolution) during the virtual experience. Since body tracking is a central requirement of VR, most of the recent HMDs are capable of tracking user's head and hands position in real-time and at high frequency, while some models include eye tracking as well. These sensors can be used for the proper positioning of the user within the VE (e.g., to update the viewpoint and stereoscopic rendering parameters), as well as to precisely determine what the user is currently looking at, but also to derive a series of additional metrics such as heartbeat and respiratory rate \cite{floris2020hmdHeartbeat}. Next-generation HMDs will directly embed dedicated sensors for monitoring such states (like the HP Reverb G2 Omnicept). 

This constant source of information can be used to acquire data that previously required to dress the user with a cumbersome set of devices and to prepare the environment for different levels of motion tracking (from a simple Microsoft Kinect to professional-grade systems such as the Vicon). Most of these capabilities are now integrated into one single device that provides all the ingredients for building an emotion recognition and elicitation system under wearable and affordable constraints. Nevertheless, HMDs can still be coupled with additional monitoring devices to increase the amount, types, and accuracy of users' bio-feedback signals for this task (e.g., by combining the full-body tracking provided by the Microsoft Kinect with the head and hands positions returned by the headset). Moreover, tools have been recently proposed to enable HMDs tracking the movements of the face\footnote{\url{https://uploadvr.com/htc-facial-tracker-quest-index/}}. In addition, VR enables simulating the context in which a user acts, and analyzing her behaviours in relation to this. Let us note that, since emotions are observable from behaviors as well, this property of VR has the potential to revolutionize the research in emotion recognition and to increase the effectiveness of this task.  

\section{The VEE-Loop}\label{sec:vee-loop}
% \textcolor{blue}{in this section, we detail the VEE loop by also explaining how it advances the state of the art}

%In this Section, we aim to define the ingredients required for the development of the VEE loop. Further, we highlight how the VEE loop and its components advance the state of the art.

%Start defining it. Then elaborate on current solutions (first subsection), and then on how our solution advances the state of the art. 

\begin{figure}[t!]
    \centering
    \includegraphics[width=0.7\linewidth]{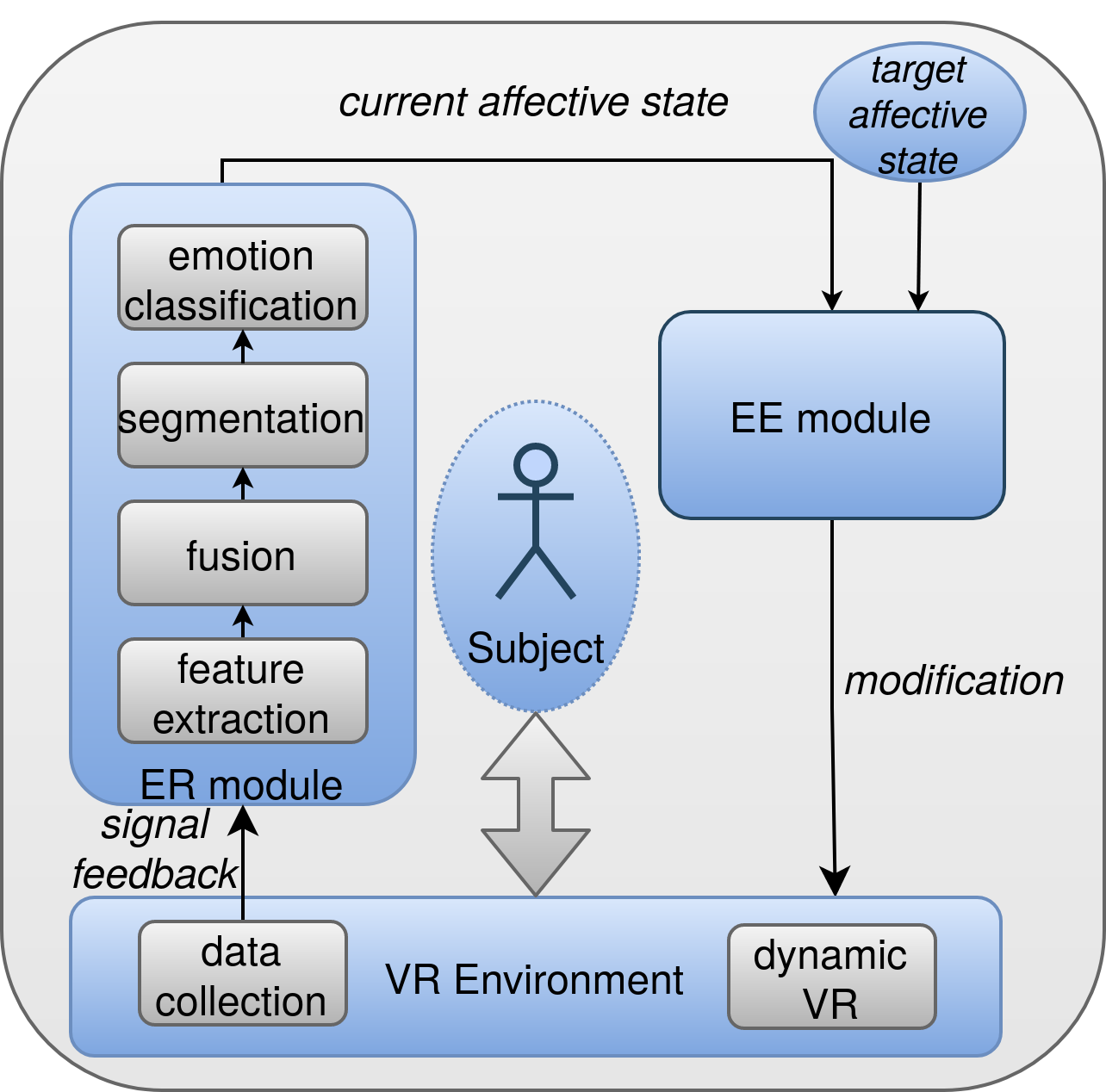}
    \caption{Architecture of the proposed VEE-Loop}
    \label{fig:veearchitecture}
\end{figure}

The VEE-loop is the realization, by means of VR technologies, of the Emotion-Recognition-Emotion-Elicitation loop described in subsection \ref{sec:loop}. More specifically, the VEE-loop is implemented by continuously monitoring the affective states of users and by adapting the VE accordingly. A modification of the VE is performed, for instance, to induce a transition from the current to the desired affective state (e.g., from fear to calm), or to evaluate the effectiveness of some emotion elicitation conditions. The VEE-loop is composed of a module for Emotion Recognition (ER module) and one for Emotion Elicitation (EE module). A detailed representation of the VEE-loop architecture is depicted in Fig. \ref{fig:veearchitecture}. From this figure, it is possible to observe that the ER module infers the emotion most likely perceived by the user by elaborating information such user's bio-feedback and the interaction between user and VE (e.g., her behaviour or the attention she pays to a particular virtual object). The emotion detected by the ER module and the emotion that the designer aims to evoke are then given in input to the EE module, which is responsible for dynamically changing the VE. We further articulate the ER and EE components in the next subsections.

%\textcolor{blue}{here we detail the VEE Loop. Figure \ref{fig:veearchitecture} shows...}

\subsection{Emotion Recognition Module}

%\textcolor{red}{in this section, we should also say that ER in VR might benefit from inputs different from users' bio-signals; indeed, the interaction with the virtual scene is per se a valuable information to obtain clues on users' emotional status.}

The ER module is responsible for inferring, from a set of multi-modal signals, the emotion that the user most likely perceives. We identify two main categories of these data: i) users' bio-feedback signals and ii) user-VE interactions. As far as user's bio-feedback are concerned (e.g., movements, vital parameters, etc.), we note that their acquisition can be performed directly with the HMD (e.g., by tracking head and eye movement), as well as with other supporting tools that do not hinder VR experience (e.g., wearable devices). Then, we also argue that users' emotions are strongly correlated with her interaction with the VE, e.g., fascination is observable from the level of attention and the time spent using some object \cite{kim2021designing}. To our knowledge, however, the problem of modeling the interaction between user and VE has never been considered in relation to the task of emotion recognition. In particular, as users' behaviours are integral aspects of their emotional status, it is essential to model the behavior of the users within the VE (e.g., which actions are taken, which types of objects are used, etc.). 

For the processing of bio-feedback and contextual data, we envision an architecture composed of the following $4$ layers: 

\begin{itemize}
    \item Feature Extraction: a set of suitable features has to be defined to capture the properties of users' bio-feedback and contextual data that are beneficial for the emotion recognition task. A quite established literature can help in the definition of features to represent a high number of bio-feedback, both handcrafted and automatically learned \cite{buccoli2016unsupervised} (e.g., acceleration of joints for body’s movements and spectrogram for voices, or learned with a deep learning approach). The definition of features to represent contextual data (e.g., interaction of the user with the VE), instead, requires a more pioneering attitude. We argue that existing features used to model the level of attention and engagement can help to define some new more emotion-oriented features.         
    \item Fusion: this layer is meant to combine data, features and algorithms to maximally exploit the information contained in the users’ data, in order to increase the generalization of the ER module; a research challenge here is to combine data of different domains (e.g., voice, heart rate and interaction with the VE) that have radically different properties, such as acquisition frequency and temporal dynamic.
    \item Segmentation: please note that, during the use of VR, users' emotions may change over time. This layer performs data segmentation, i.e., it segments the stream into portions of signals that are coherent with respect to the particular emotion they carry. This is a remarkable difference with respect to existing studies on emotion recognition, which generally assume that observed data is associated with a single emotional content.
    \item Emotion Classification: finally, this layer infers the emotion that the obtained segmented data most likely carries. Note that emotions can be represented either using a set of classes (e.g., joy, fear, etc.), or using a dimensional approach (e.g., arousal, valence, dominance \cite{mehrabian1996pleasure}). Based on the type of representation that is chosen, this layer performs a supervised learning task, either a classification or a regression.  
\end{itemize}

%To summarize, the architecture of the ER module infers emotions from a continuous stream of users’ generated data (and not, as commonly done in ER research, from standalone data, i.e., signals associated with a single emotion), and works both on single-mode (e.g., on users’ movements only) and multi-mode manner (e.g., on a combination of users’ movements and voice).

\subsection{Emotion Elicitation Module}

The Emotion Elicitation (EE) module outputs a modified VE based on the following input: i) a representation of the current VE, ii) the emotion detected by the ER module and the iii) the emotion that designer aims to evoke. Firstly, the EE module computes a measure of distance between the emotion intended by the designer and that recognized by the ER module. Then, based on this distance, the properties of the current VE are appropriately modified (e.g., the colour of a given virtual object is changed).  

The main research questions that are still pending here are how to measure the distance between emotions, and how to modify the VE accordingly. We note that, in order to compute this distance, emotions should be better described using a dimensional representation (e.g., in the valance/arousal plane), which allows quantifying the difference among them. Then, the difficulty in modifying a content to elicit emotions is a well-known problem, in particular when advanced interactive media, such as VR, are considered \cite{riva2007affective}. In our view, the first step to tackle this problem is the definition of a representation of the VE that includes, for instance, positions, semantic (i.e., functional role) and the sensory qualities (e.g., shape and size) of the most salient virtual objects. The second step is the definition of a model that, as a function of the detected emotion, its distance with the target emotion, and the representation of the current VE, returns an indication on how the VE should be modified. For instance, if target emotion is joy, but the one detected by the ER module is sadness, the colors of objects should be tuned to be more warmed. In our view, this task is still very complex to be automatized, and would require the manual tuning of VE's characteristics. However, in case the VEE-loop became a tool of common use, it would function also as a tool for data collection. Specifically, the tuning choices of designers, along with the emotional reactions of users, might be collected and then used to train automatic systems (e.g., machine learning algorithms). 

\section{Advancing the State of the Art}\label{sec:soa}

Early prototypes of the VEE-loop are present in the literature. For instance, in \cite{i2018toward} an architecture to perform users’ emotion-driven generation of a VE is proposed and validated in the context of mental health treatment. Such architecture is designed to detect users’ emotions from the analysis of multiple types of bio-feedback (similarly to our ER module) and, accordingly, to generate a VE to stabilize them, e.g., to induce calm (similarly to our EE module). Whilst this existing approach is quite similar, in principle, to our idea of VEE-loop, it has the main drawback of not considering complex models of interaction between users and VE that, in our view, are essential to realize an instrument suitable for emotion-driven product design (for instance, in Ref. \cite{i2018toward} the generated VE is a simple maze). Another work that investigates the use of VR as a tool to perform ER and EE can be found in Ref. \cite{marin2020modelling}, where ER is performed using a very simple machine learning algorithm that works on users' electroencephalograms, while EE is implemented using static VEs. Instead, in order to be a suitable tool for emotion-driven product design, the VEE-loop will consider a vast array of heterogeneous bio-feedback, complex models of interaction between users and the VE, dynamic VEs and more advanced machine learning algorithms.

Most of the research on ER is done on single-mode and standalone data (see the recent survey \cite{saxena2020emotion}), which carry acted and exaggerated emotions. Instead, the proposed framework allows considering streams of multi-mode data (which introduce the challenge of identifying the onset and end of emotions) and exploiting the sense of presence typical of VR experience to induce (and then, recognize) more spontaneous emotions \cite{susindar2019feeling}. %We advance the state of the art on this field (which is currently far from being mature) by performing massive experiments across various areas, such as education (as done in Ref. \cite{dosseville2012music} using other elicitation media), advertising (to understand what creates favourable and unique brand associations \cite{keller2001building} and enhances consumer-brand relationships \cite{brodie2013consumer}) and interior design (to understand which factors reinforce people’s wellbeing \cite{de2016emotions,de2016oltre}).

%\textcolor{blue}{we might add a table to show the differences wrt the state of the art}

Refs. \cite{alaniz2019emotional} and \cite{kim2021designing} describe methodological frameworks that designer can follow to develop products with emotional intentions. However, as far as we know, we are the first to propose a technological solution that can help in the design of emotion-driven products. We also remark that the proposed VEE-loop can also be used to dynamically modify virtual products based on users' emotions (e.g., a service of remote schooling). In this respect, similar previous work (that, however, do not make use of VR technologies) are \cite{polignano2021towards,mariappan2012facefetch,sindhu2021emotion,rumiantcev2020emotion} and \cite{liu2013eeg}, which propose systems for emotion-driven recommendation and advertising, respectively. Similarly, a digital system that adapts the characteristics of a service based on users' emotions is proposed in \cite{condori2017happyness}. Then, Ref. \cite{hossain2015audio} proposes a gaming framework that changes the characteristics of the game based on users' emotions. Finally, Ref. \cite{munoz2018emotion} describes the realization of a smart office in which sensory features (e.g., light in the office) and tasks assigned to users are changed to regulate their emotions.

\section{Impact and Application}\label{sec:impact_application}
In this Section, we describe the main practical applications of the VEE-loop, as well as its potential impact across several areas.

\subsection{Areas of application}
\label{sec:applicative_areas}

The VEE-loop is a versatile tool that opens the doors to a wide spectrum of applications. In particular, we identify the following three main areas of applications: 1) product design, 2) virtual service delivery and 3) research in emotion recognition and elicitation. 

In product design, the VEE-loop can be used to validate the capability of a product to fulfill its emotional requirements (i.e., to check the consistency between intended and perceived emotions) before the actual and expensive tangible production. In fact, by exploiting the sense of presence given by VR, the emotions experienced by the users are guaranteed to be as much similar as possible to the real ones. Hence, users can try the virtual counterparts of the products under development, and provide the designers with implicit feedback about the goodness of their functional and stylistic choices. Note also that, being the VEE-loop a digital tool, this validation can easily involve a higher number of users with respect to traditional on site experiments, therefore increasing their validity. Beside improving the product, the received feedback also helps designers to better study their customers at the emotional level and to understand which factors reinforce brand identification \cite{de2016emotions,de2016oltre}.

Moreover, services that are delivered using digital channels can benefit from the use of the VEE-loop. We refer, in particular, to services in the education field, where having the information on users’ emotional states is highly beneficial \cite{ortigosa2014sentiment}, but unavailable for some reasons (e.g., in remote schooling during the Covid-19 pandemic), or in human-machine interaction, where the required equipment is too much expensive or dangerous (e.g., in the training of practitioners in industry). In this type of services, which are delivered in real-time, the VEE-loop is implemented to adapt the VE to the current emotional status of the users, e.g., to calm them down when they are anxious. This can be done by modifying the sensory qualities of the virtual objects (as also done in product design), as well as by adapting the learning tasks to enhance users' experience. Moreover, the VEE-loop can facilitate the transition towards an increasingly-digitized society, where a number of services can be modified according to users' emotions. In theatrical exhibitions, for instance, the VEE-loop can be exploited to better understand the relation between emotions and acting.

Finally, the VEE-loop has the potential to advance the state of the art on the growing fields of emotion recognition and elicitation. As for the former, the VEE-loop can help to enhance current models for emotion recognition, e.g., by including the context embodied in the VE and users' behaviours. As for the latter, the VEE-loop can be exploited in many different research areas to better study the effectiveness of elicitation conditions (e.g., in marketing, interior design, etc.).

\subsection{Potential Impact}

% We advance the state of the art on this field (which is currently far from being mature) by performing massive experiments
In light of the numerous potential applications described before, our proposed solution can potentially benefit various dimensions of our society, as detailed in the following.

\paragraph{Economical Impact}
The VEE loop finds applications in a countless number of industrial sectors, while providing potential economical advantages both in the production and in the marketing phases. 
% which will bring research-oriented and applicative collaborations. 
For example, it can be used by experts in advertising to understand what reinforces unique brand association, or by designers to evaluate users’ emotional response to the characteristics of a product before its tangible development. This allows designers to take more informed decisions, therefore, reducing the risks (and associated costs) of creating unsuccessful products and services.

\paragraph{Social Impact}
The VEE loop can help to deliver more empathetic services using VR, therefore bringing a high social impact across many different areas (e.g., remote schooling). Potential applications can also target the treatment of pathologies characterized by disorders on the emotional sphere \cite{i2018toward} and collective training in emergency situations.

\paragraph{Environmental Impact}
The VEE loop integrates emotional aspects into services delivered remotely, therefore, increasing their adoption. This has the potential of improving remote working and practices, thus, limiting unnecessary travels, and, in turn, reducing the emissions produced by means of transport.

\paragraph{Research Impact}
Our vision contributes to the research on ER and EE, and provides a tool that researchers can readily use in the studies relative to these fields. %Moreover, the obtained methodological results can be made available to students in several courses, such as those on product and graphic design, interior design and intelligent systems and services. 
The VEE loop is a novel and timely solution that can be a potential cornerstone in many different projects (from the research-oriented to the more applicative ones), therefore, enabling transversal collaborations between academy and industry.

\paragraph{Cultural Impact}
The VEE loop enables avant-garde cultural events delivered with VR. For instance, stylistic choices of a cultural event (e.g., in theatrical representations) can be modified according to the emotional response of the audience (even if attending remotely) in real-time and in an economically-sustainable manner. This asset can find application in several cultural scenarios, e.g., theatre, virtual city trip and museum virtual tours.

%\subsubsection{Illustrative Example}

%let's describe an example here. 

\section{Discussion}\label{sec:challenges}

In this paper, we have identified the iteration of emotion elicitation and recognition phases as a desirable property of an emotion-driven design strategy. We have then provided arguments to support the idea of using VR to realize this iteration, that we have referred to as the Virtual-Reality-Based Emotion-Elicitation-and-Recognition loop (VEE-loop). In brief, VR allows creating virtual yet very realistic environments that designers can flexibly modify to induce specific emotional reactions. Indeed, VR inherits all the benefits of digital technologies (e.g., flexible and controlled content modification), without sacrificing the realism of the experience. All these aspects render the VEE-loop a promising methodological and technological framework that can benefit various areas, such as product design, virtual service delivery and, more in general, the research in emotion recognition and elicitation. However, several issues still need to be properly tackled before its adoption. In the following, we summarize the main characteristics of the VEE-loop, and we discuss corresponding opportunities and challenges. % To facilitate the reading, we discuss these points following the %. For an easier reading, we frame the discussion according to the main application areas identified in subsection \ref{sec:applicative_areas}, namely product design, virtual service delivery and research in emotion recognition and elicitation. 

Our most important claim is that the VEE-loop can help designers developing products capable to induce some specific emotion on their users. The flexible content adaptation and sense of immersion guaranteed by VR are just a couple of reasons that support this claim. In fact, VR allows experimenting a large number of virtual products' characteristics, which can evoke emotions similarly to their real counterpart. However, the right emotion elicitation conditions are likely to be found after a number of iterations of the VEE-loop. The main problem of this iterative approach is that the emotions that users perceive are not only influenced by the current virtual content, but also by the number of iteration itself (e.g., stress and boredom might arise after a long session of experiments). This issue needs to be correctly tackled to not hinder the validity of the performed experiments. When the VEE-loop is used in the design phase, a possible approach consists in limiting the duration of the experiment to a certain amount of time, and to validate the effectiveness of emotion elicitation conditions through statistical analysis (e.g., which emotions have been perceived by the majority of the users involved in the experiments). As for experiments done to understand the emotional reaction of a specific user, instead, a possible strategy is to alternate VEs that carry an emotional content with VEs that are emotionally neutral, so to bring the user back to her normal conditions.  

%\textcolor{red}{Desmet (2002) states	 that due	to	emotional	reactions being personal and different for each individual,	foretelling	or even influencing the	emotional impact of	a design	to	induce	targeted	emotional	responses	is	difficult. Let us elaborate on this, saying that this is a challenge for our proposed system as well.}

More generally, VR allows simulating the context in which a product is used, e.g., the surrounding environment and the after-use experience. In light of this, the VEE-loop becomes even a more powerful design instrument, as designers are free to experiment a higher number of emotion elicitation conditions, and emotion recognition exploits a richer set of observations, e.g., both bio-feedback signals and users' behaviour. Two open research challenges here are i) how to define suitable experiments to gain relevant insights from users' behaviours, which are an essential aspect of their emotions (e.g., how users interact with the VE, which elements they observe, etc.) and ii) how to automatize the dynamic modification of the VE.

Then, the high level of immersion guaranteed by VR lead users to perceive more spontaneous emotions that, for this reason, are a more valuable feedback for the designers, but also more difficult to identify. Indeed, most of the research on emotion recognition is based on the analysis of emotions that are voluntarily exaggerated and that, for this reason, are also easier to recognize. Finally, emotions must be estimated from a stream of signals and not, as generally done in previous work, from data that are assumed to carry a single emotion. Therefore, a segmentation process is required to identify the instants of transitions between two different affective states, before their actual classification. To our knowledge, the problem of segmentation is extensively considered in the action recognition task, but quite unexplored in the emotion recognition one. 

To conclude, the VEE-loop represents a promising methodological and technological framework that can be exploited to effectively design emotion-driven products, provided that several issues (both at the technological and methodological sides) are correctly handled.  

\bibliography{biblio}

\begin{thebibliography}{10}
\providecommand{\url}[1]{#1}
\csname url@samestyle\endcsname
\providecommand{\newblock}{\relax}
\providecommand{\bibinfo}[2]{#2}
\providecommand{\BIBentrySTDinterwordspacing}{\spaceskip=0pt\relax}
\providecommand{\BIBentryALTinterwordstretchfactor}{4}
\providecommand{\BIBentryALTinterwordspacing}{\spaceskip=\fontdimen2\font plus
\BIBentryALTinterwordstretchfactor\fontdimen3\font minus
  \fontdimen4\font\relax}
\providecommand{\BIBforeignlanguage}[2]{{%
\expandafter\ifx\csname l@#1\endcsname\relax
\typeout{** WARNING: IEEEtran.bst: No hyphenation pattern has been}%
\typeout{** loaded for the language `#1'. Using the pattern for}%
\typeout{** the default language instead.}%
\else
\language=\csname l@#1\endcsname
\fi
#2}}
\providecommand{\BIBdecl}{\relax}
\BIBdecl

\bibitem{alaniz2019emotional}
T.~Alaniz and S.~Biazzo, ``Emotional design: the development of a process to
  envision emotion-centric new product ideas,'' \emph{Procedia Computer
  Science}, vol. 158, pp. 474--484, 2019.

\bibitem{zaltman2003subconscious}
G.~Zaltman, ``The subconscious mind of the consumer (and how to reach it),''
  \emph{Harvard Business School. Working Knowledge. Obtenido de http://hbswk.
  hbs. edu/item/3246. html}, 2003.

\bibitem{wrigley2019affected}
C.~Wrigley and K.~Straker, \emph{Affected: Emotionally engaging customers in
  the digital age}.\hskip 1em plus 0.5em minus 0.4em\relax John Wiley \& Sons,
  2019.

\bibitem{riva2007affective}
G.~Riva, F.~Mantovani, C.~S. Capideville, A.~Preziosa, F.~Morganti, D.~Villani,
  A.~Gaggioli, C.~Botella, and M.~Alca{\~n}iz, ``Affective interactions using
  virtual reality: the link between presence and emotions,''
  \emph{CyberPsychology \& Behavior}, vol.~10, no.~1, pp. 45--56, 2007.

\bibitem{ortigosa2014sentiment}
A.~Ortigosa, J.~M. Mart{\'\i}n, and R.~M. Carro, ``Sentiment analysis in
  facebook and its application to e-learning,'' \emph{Computers in human
  behavior}, vol.~31, pp. 527--541, 2014.

\bibitem{liu2013eeg}
Y.~Liu, O.~Sourina, and M.~R. Hafiyyandi, ``Eeg-based emotion-adaptive
  advertising,'' in \emph{2013 Humaine Association Conference on Affective
  Computing and Intelligent Interaction}.\hskip 1em plus 0.5em minus
  0.4em\relax IEEE, 2013, pp. 843--848.

\bibitem{desmet2002designing}
P.~Desmet, ``Designing emotions,'' 2002.

\bibitem{sinek2009start}
S.~Sinek, \emph{Start with why: How great leaders inspire everyone to take
  action}.\hskip 1em plus 0.5em minus 0.4em\relax Penguin, 2009.

\bibitem{frijda1986emotions}
N.~H. Frijda \emph{et~al.}, \emph{The emotions}.\hskip 1em plus 0.5em minus
  0.4em\relax Cambridge University Press, 1986.

\bibitem{desmet2016emotion}
P.~M. Desmet, S.~F. Fokkinga, D.~Ozkaramanli, and J.~Yoon, ``Emotion-driven
  product design,'' in \emph{Emotion Measurement}.\hskip 1em plus 0.5em minus
  0.4em\relax Elsevier, 2016, pp. 405--426.

\bibitem{kim2021designing}
C.~Kim, J.~Yoon, P.~Desmet, and A.~Pohlmeyer, ``Designing for positive
  emotions: Issues and emerging research directions,'' 2021.

\bibitem{yoon2016feeling}
J.~Yoon, A.~E. Pohlmeyer, and P.~Desmet, ``When'feeling good'is not good
  enough: Seven key opportunities for emotional granularity in product
  development,'' \emph{International Journal of Design}, vol.~10, no.~3, pp.
  1--15, 2016.

\bibitem{diemer2015impactPerception}
\BIBentryALTinterwordspacing
J.~Diemer, G.~W. Alpers, H.~M. Peperkorn, Y.~Shiban, and A.~Mühlberger, ``The
  impact of perception and presence on emotional reactions: a review of
  research in virtual reality,'' \emph{Frontiers in Psychology}, vol.~6, p.~26,
  2015. [Online]. Available:
  \url{https://www.frontiersin.org/article/10.3389/fpsyg.2015.00026}
\BIBentrySTDinterwordspacing

\bibitem{morales2019realVsVRemotion}
\BIBentryALTinterwordspacing
J.~Marín-Morales, J.~L. Higuera-Trujillo, A.~Greco, J.~Guixeres, C.~Llinares,
  C.~Gentili, E.~P. Scilingo, M.~Alcañiz, and G.~Valenza, ``Real vs.
  immersive-virtual emotional experience: Analysis of psycho-physiological
  patterns in a free exploration of an art museum,'' \emph{PLOS ONE}, vol.~14,
  no.~10, pp. 1--24, 10 2019. [Online]. Available:
  \url{https://doi.org/10.1371/journal.pone.0223881}
\BIBentrySTDinterwordspacing

\bibitem{babu2018vrLearningRate}
S.~{K. Babu}, S.~{Krishna}, U.~{R.}, and R.~R. {Bhavani}, ``Virtual reality
  learning environments for vocational education: A comparison study with
  conventional instructional media on knowledge retention.'' in \emph{2018 IEEE
  18th International Conference on Advanced Learning Technologies (ICALT)},
  2018, pp. 385--389.

\bibitem{floris2020hmdHeartbeat}
\BIBentryALTinterwordspacing
C.~Floris, S.~Solbiati, F.~Landreani, G.~Damato, B.~Lenzi, V.~Megale, and E.~G.
  Caiani, ``Feasibility of heart rate and respiratory rate estimation by
  inertial sensors embedded in a virtual reality headset,'' \emph{Sensors},
  vol.~20, no.~24, 2020. [Online]. Available:
  \url{https://www.mdpi.com/1424-8220/20/24/7168}
\BIBentrySTDinterwordspacing

\bibitem{buccoli2016unsupervised}
M.~Buccoli, M.~Zanoni, A.~Sarti, S.~Tubaro, and D.~Andreoletti, ``Unsupervised
  feature learning for music structural analysis,'' in \emph{2016 24th European
  Signal Processing Conference (EUSIPCO)}.\hskip 1em plus 0.5em minus
  0.4em\relax IEEE, 2016, pp. 993--997.

\bibitem{mehrabian1996pleasure}
A.~Mehrabian, ``Pleasure-arousal-dominance: A general framework for describing
  and measuring individual differences in temperament,'' \emph{Current
  Psychology}, vol.~14, no.~4, pp. 261--292, 1996.

\bibitem{i2018toward}
S.~B. i~Badia, L.~V. Quintero, M.~S. Cameir{\~a}o, A.~Chirico, S.~Triberti,
  P.~Cipresso, and A.~Gaggioli, ``Toward emotionally adaptive virtual reality
  for mental health applications,'' \emph{IEEE journal of biomedical and health
  informatics}, vol.~23, no.~5, pp. 1877--1887, 2018.

\bibitem{marin2020modelling}
J.~Mar{\'\i}n~Morales, ``Modelling human emotions using immersive virtual
  reality, physiological signals and behavioural responses,'' Ph.D.
  dissertation, 2020.

\bibitem{saxena2020emotion}
A.~Saxena, A.~Khanna, and D.~Gupta, ``Emotion recognition and detection
  methods: A comprehensive survey,'' \emph{Journal of Artificial Intelligence
  and Systems}, vol.~2, no.~1, pp. 53--79, 2020.

\bibitem{susindar2019feeling}
S.~Susindar, M.~Sadeghi, L.~Huntington, A.~Singer, and T.~K. Ferris, ``The
  feeling is real: Emotion elicitation in virtual reality,'' in
  \emph{Proceedings of the Human Factors and Ergonomics Society Annual
  Meeting}, vol.~63, no.~1.\hskip 1em plus 0.5em minus 0.4em\relax SAGE
  Publications Sage CA: Los Angeles, CA, 2019, pp. 252--256.

\bibitem{polignano2021towards}
M.~Polignano, F.~Narducci, M.~de~Gemmis, and G.~Semeraro, ``Towards
  emotion-aware recommender systems: an affective coherence model based on
  emotion-driven behaviors,'' \emph{Expert Systems with Applications}, vol.
  170, p. 114382, 2021.

\bibitem{mariappan2012facefetch}
M.~B. Mariappan, M.~Suk, and B.~Prabhakaran, ``Facefetch: A user emotion driven
  multimedia content recommendation system based on facial expression
  recognition,'' in \emph{2012 IEEE International Symposium on
  Multimedia}.\hskip 1em plus 0.5em minus 0.4em\relax IEEE, 2012, pp. 84--87.

\bibitem{sindhu2021emotion}
N.~Sindhu, S.~Jerritta, and R.~Anjali, ``Emotion driven mood enhancing
  multimedia recommendation system using physiological signal,'' in \emph{IOP
  Conference Series: Materials Science and Engineering}, vol. 1070,
  no.~1.\hskip 1em plus 0.5em minus 0.4em\relax IOP Publishing, 2021, p.
  012070.

\bibitem{rumiantcev2020emotion}
M.~Rumiantcev and O.~Khriyenko, ``Emotion based music recommendation system,''
  in \emph{Proceedings of Conference of Open Innovations Association
  FRUCT}.\hskip 1em plus 0.5em minus 0.4em\relax Fruct Oy, 2020.

\bibitem{condori2017happyness}
N.~Condori-Fernandez, ``Happyness: an emotion-aware qos assurance framework for
  enhancing user experience,'' in \emph{2017 IEEE/ACM 39th International
  Conference on Software Engineering Companion (ICSE-C)}.\hskip 1em plus 0.5em
  minus 0.4em\relax IEEE, 2017, pp. 235--237.

\bibitem{hossain2015audio}
M.~S. Hossain, G.~Muhammad, B.~Song, M.~M. Hassan, A.~Alelaiwi, and A.~Alamri,
  ``Audio--visual emotion-aware cloud gaming framework,'' \emph{IEEE
  Transactions on Circuits and Systems for Video Technology}, vol.~25, no.~12,
  pp. 2105--2118, 2015.

\bibitem{munoz2018emotion}
S.~Munoz, O.~Araque, J.~F. S{\'a}nchez-Rada, and C.~A. Iglesias, ``An emotion
  aware task automation architecture based on semantic technologies for smart
  offices,'' \emph{Sensors}, vol.~18, no.~5, p. 1499, 2018.

\bibitem{de2016emotions}
V.~De~Luca, ``Emotions-based interactions: Design challenges for increasing
  well-being,'' 2016.

\bibitem{de2016oltre}
------, ``Oltre l’interfaccia: emozioni e design dell’interazione per il
  benessere,'' \emph{MD Journal}, vol.~1, no.~1, pp. 106--119, 2016.

\end{thebibliography}
\bibliographystyle{IEEEtran}

\end{document}